\documentclass[nonacm,sigconf]{acmart}

\usepackage{multirow}
\newcommand{\techreportURL}{https://research.engr.oregonstate.edu/idea/sites/research.engr.oregonstate.edu.idea/files/scalable_schema_mapping-technical_report.pdf}

\theoremstyle{definition}
\newtheorem{example}{Example}

\begin{document}

\title{Towards Scalable Schema Mapping using Large Language Models}

\author{Christopher Buss}
\authornote{Both authors contributed equally to this research.}
\email{bussch@oregonstate.edu}
\affiliation{
  \institution{Oregon State University}
  \city{Corvallis}
  \state{Oregon}
  \country{USA}
}

\author{Mahdis Safari}
\authornotemark[1]
\email{safarim@oregonstate.edu}
\affiliation{
  \institution{Oregon State University}
  \city{Corvallis}
  \state{Oregon}
  \country{USA}
}

\author{Arash Termehchy}
\email{termehca@oregonstate.edu}
\affiliation{
  \institution{Oregon State University}
  \city{Corvallis}
  \state{Oregon}
  \country{USA}
}

\author{Stefan Lee}
\email{leestef@oregonstate.edu}
\affiliation{
  \institution{Oregon State University}
  \city{Corvallis}
  \state{Oregon}
  \country{USA}
}

\author{David Maier}
\email{maier@pdx.edu}
\affiliation{
  \institution{Portland State University}
  \city{Portland}
  \state{Oregon}
  \country{USA}
}

\renewcommand{\shortauthors}{Buss et al.}

\begin{abstract}
The growing need to integrate information from a large number of diverse sources poses significant scalability challenges for data integration systems. These systems often rely on manually written schema mappings, which are complex, source-specific, and costly to maintain as sources evolve. While recent advances suggest that large language models (LLMs) can assist in automating schema matching by leveraging both structural and natural language cues, key challenges remain. In this paper, we identify three core issues with using LLMs for schema mapping: (1) \textbf{inconsistent outputs} due to sensitivity to input phrasing and structure, which we propose methods to address through sampling and aggregation techniques; (2) the need for \textbf{more expressive mappings} (e.g., GLaV), which strain the limited context windows of LLMs; and (3) the \textbf{computational cost} of repeated LLM calls, which we propose to mitigate through strategies like data type prefiltering.
\end{abstract}

\maketitle

\section{Introduction}
\label{sec:introduction}

There is a recognized need to collect and connect information from a variety of data sources \cite{Doan:2012:PDI:2401764,DBLP:conf/pods/GolshanHMT17,NSF_smartHealth}.
As an example, we have recently worked in a large-scale NIH-funded project to augment the information of biomedical entities by querying other biomedical data sources \cite{Wood2021.10.17.464747}.
The main focus of this project is to {\it repurpose} current drugs to treat or mitigate the symptoms of diseases for which there is insufficient time or resources to develop effective treatments (e.g., new or rare diseases) \cite{Drug-repositioning}.
To support such a system, its developers must find and combine a patchwork of data sources to get a full picture of a drug (e.g., clinical trials, research literature, and adverse effects). 
Collecting all this information is resource-intensive and can be a barrier to important discoveries.

Data integration systems are complex and often ingest data from a large number of diverse sources.
For each source, programmers must manually write mappings that reconcile structural differences.
Writing the correct mappings often requires understanding the semantics of the source. 
Thus, programmers must cross-reference natural-language descriptions (e.g., database documentation), the logical model, and the actual representation of data.

Due to the complexity of mappings, the large number of sources, and the fact that sources evolve over time, integration systems have major scalability problems. 
Often, mappings are source-specific and cannot be reused.
Systems not only become more complex as mappings are added, but with each new source, there is a higher probability that any one source will change, disrupting the system until developers repair the affected mappings. 
This results in high maintenance costs.
To contend with the growing number of sources, we must develop new tools to reduce the manual effort required to build and maintain data integration systems.

Due to the complicated nature of data integration, it requires human-in-the-loop approaches.
Many older works have focused on rule-based tools \cite{popa2002translating, Coma1, bonifati2008schema}, which lack semantic understanding, limiting their usefulness.
More recent work has proposed training models to expand their ability beyond basic rules \cite{zhang2021smat}.
However,  doing so requires preparing labeled data specific to certain domains.

Recent studies have explored how large language models (LLMs) can be used to generate schema mappings \cite{ReMatch, EnhancingBiomedicalSM, MatchMaker, GRAM, KcMF, TaDa, huang2150transform}.
Since LLMs can incorporate a wide range of supporting information, including schema metadata and natural language descriptions, they are especially suited to generating mappings.

In this paper we describe three challenges with using LLMs for generating schema mappings.
\begin{itemize}
    \item \textbf{Inconsistent Outputs}: LLMs are highly sensitive to input phrasing and structure, leading to unpredictable and diverse results.
    We propose techniques for sampling and combining multiple outputs, increasing our mapping coverage and providing effective ways to filter out unlikely mappings.
    We provide preliminary results that illustrate the effectiveness of our methods.
    \item \textbf{GLaV with Limited Contexts}: Existing LLM-based methods focus on limited mapping types, which are often insufficient for many real-world integration scenarios.
    Towards supporting more integration systems, we consider the challenges associated with generating more expressive mappings (GLaV). Supporting such mappings would increase complexity, requiring more sophisticated representations and careful design to avoid overwhelming the LLM’s context.
    \item \textbf{Challenge: Efficient Prompting}: LLM-based schema mapping is computationally expensive due to repeated model calls, especially with large datasets.
    This, in large part, is thanks to its large input, which only becomes more difficult to manage as we consider more expressive mappings.
\end{itemize}
\section{Schema Mapping Assistant}
\label{sec:schema_mapping}

\begin{figure}
	\centering
        \includegraphics[width=1\linewidth]{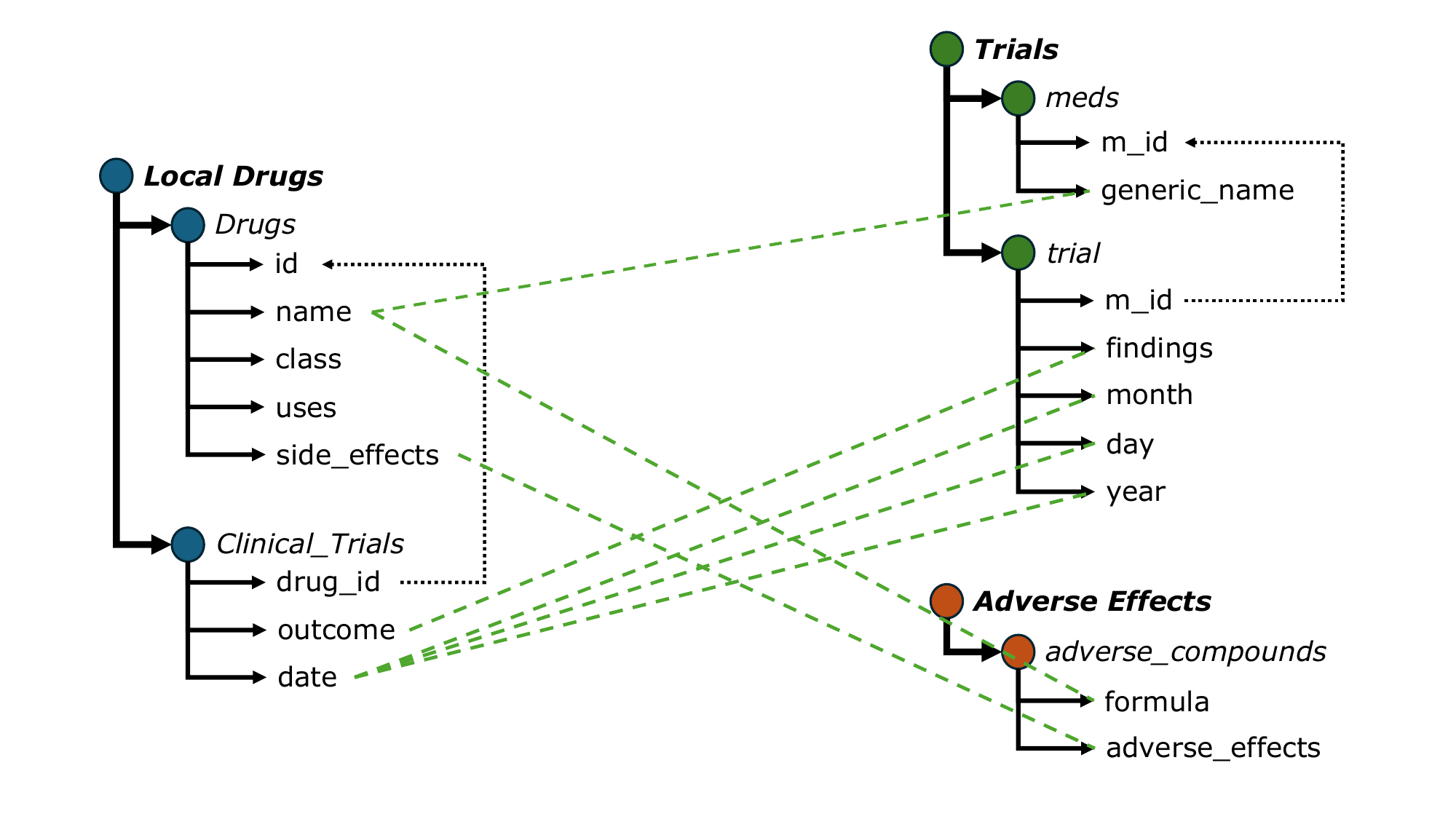}
    \caption{
Example integration scenario with a target (left) and two sources (right). 
Green dashed lines represent semantic correspondences between attributes (schema alignment). 
Dotted represent inter-schema references (foreign keys).
}
    \label{fig:connecting}

    \Description{
The figure shows multiple sources with different column names for the same drug-related information. It highlights the need to align local drug attributes with external ones to create a unified view. Lines between tables indicate corresponding attributes, and arrows show foreign key references between schemas.
}

\end{figure}

Our discussion is inspired by tools \cite{bonifati2008schema, popa2002translating, Coma1}, meant to assist the user in the schema mapping process.
Generally speaking, these systems take the full set of possible mappings and filter them down to a candidate set. 
The candidate set is then shown to the user so they may verify them, selecting which mappings to implement.
This work focuses on mappings between relational databases, but we assert that the challenges highlighted here apply to schema mapping broadly, regardless of the logical models used.

It is expected that users will need to verify the output mappings as the underlying model used to generate them will not know the user's latent intent. 
In essence, there is nearly always expectations (e.g., business rules not represented within the schema) for which the underlying model is not privy to.

\subsection{Preliminaries}
\paragraph{Data Integration System}
Following previous works \cite{lenzerini2002data}, we describe a data integration system $\mathcal{I}$ as a triple $\langle\mathcal{G}, \mathcal{S}, \mathcal{M}\rangle$, where
\begin{itemize}
    \item $\mathcal{G}$ is the \textit{target schema}, which describes a unified view of sources.
    \item $\mathcal{S}$ is the \textit{source schema}, which specifies the structure of the sources to integrate. For the sake of definitional simplicity, we do not distinguish between different sources; instead, we consider $\mathcal{S}$ to simply be the union of all source schemas. 
    \item $\mathcal{M}$ is the mapping between $\mathcal{G}$ and $\mathcal{S}$, constituted by a set of \textit{rules}, each describing how a subset of $\mathcal{G}$ semantically corresponds to a subset of $\mathcal{S}$.
\end{itemize}
Both $\mathcal{S}$ and $\mathcal{G}$ contain relations $S_1, S_2, ... S_n$ and $G_1, G_2, ... G_m$ respectively.
In turn, each relation contains a set of attributes denoted as $attr(S_i)=\{s_1, s_2, ..., s_l\}$ and $attr(G_j)=\{g_1, g_2, ..., g_k\}$.

\begin{example}
Figure~\ref{fig:connecting} represents an integration scenario within the drug domain.
Specifically, to better understand whether certain drugs can be used to treat certain rare diseases, we want to integrate information about each drug's clinical trials and adverse affects, and likely many other sources not picture here.
In our running example, we focus on integrating information from the Trials source:
given $\mathcal{S} = \{meds, trial\}$ and $\mathcal{G} = \{Drugs, Clinic\_Trials\}$, we must define a mapping ($\mathcal{M}$).
\end{example}

\paragraph{Mapping Rules (st-tgds)}
We formally express rules as Source-to-Target Tuple-Generating Dependencies (st-tgds) \cite{fagin2005data},
\[
\forall \vec{x} \; \big( \phi(\vec{x}) \rightarrow \exists \vec{y} \; \psi(\vec{x}, \vec{y}) \big)
\]
where
\begin{itemize}
    \item $\phi(\vec{x})$ is a conjunction of atoms over the source $\mathcal{S}$.
    \item $\psi(\vec{x}, \vec{y})$ is a conjunction of atoms over the target $\mathcal{G}$.
    \item $\vec{x}$ are universally quantified variables.
    \item $\vec{y}$ are existentially quantified variables.
\end{itemize}
Both $\phi(\vec{x})$ and $ \psi(\vec{x}, \vec{y})$ may include additional predicates (e.g., for filtering tuples).
In essence, each rule asserts a pattern over the source, that, if matched, generates tuples adhering to the corresponding pattern in the target.

\begin{example}
Continuing our example in~\ref{fig:connecting}, we indicate which tuples in Trials should trigger the generation of tuples in Local Drugs.
Further, we indicate which attributes within the new target tuples should be populated, and how that population is determined by the attributes within the triggering source tuples.
\begin{equation}
\begin{aligned}
\forall i, g, f, m, &\ d, y \; \big(
  \text{meds}(i, g) \land \text{trial}(i, f, m, d, y), y>1990 \\
  &\rightarrow \exists x_1, x_2, x_3, x_4 \; \big(
    \text{Drugs}(x_1, \tau_1[g], x_2, x_3, x_4) \land {} \\
  &\hspace{4.0em} \text{Clinical\_Trials}(x_1, f, \tau_2[m, d, y])
  \big)
\big)
\end{aligned}
\label{eq:glav_tgd}
\end{equation}
Target attributes are populated based on their semantic counterparts in the source.
In some instances, value-level transformations are necessary, such as translating a drug's generic name to its brand name ($\tau_1$) and concatenating date-parts ($\tau_2$).
However, we make a distinction between value-level transformations and schema-level transformations.
This work focuses on the latter, though, in the long-term, we foresee it being useful, especially for complex value-level transformations, to tackle the former problem as a separate step from that of schema-level transformations.
\end{example}

\paragraph{Referential Dependencies}
Often, rules contain \textit{referential dependencies} which condition the existence of rows in one relation upon the existence of join-able rows in another.
In the case of rule~\ref{eq:glav_tgd}, we specify referential dependencies over both the source and target.
Over the source, the shared variable $i$ in $meds$ and $trial$ forces tuples from these relations to fall within the same equi join on \textit{m\_id}.
Over the target, the shared variable $x_1$ implies the creation of a surrogate key for each answer within the source, ensuring that rows in $Drugs$ are connected to their corresponding rows in $Clinical\_Trials$.
Importantly, referential dependencies are expressed via the presence of the same variable ($i$ and $x_1$) within \textit{multiple} relational predicates on the left-hand side ($meds$ and $trials$) and right-hand side ($Drugs$ and $Clinical\_Trials$) of the \textit{same} rule.

\subsection{Rule Expressiveness}
\label{subsec:expressivness}
Rules are commonly divided into three classes, each of which is defined by the number of relational predicates allowed over the source and target.
Formally, these classes are called Global-as-View (GaV), Local-as-View (LaV), and Global-Local-as-View (GLaV).
We refer interested readers to \cite{lenzerini2002data} for a more detailed comparison of these three classes.
For the sake of our exposition, we simply make the distinction between those classes that limit either side of a rule to one-and-only-one relational predicate (i.e., GaV and LaV) and the class that does not (i.e., GLaV).
Henceforth refer to the former class as \textit{limited referential dependencies (LRD)} and the latter class as \textit{full referential dependencies (FRD)}.
The rule written in Equation~\ref{eq:glav_tgd} falls strictly within the FRD class as more than one relational predicate appears on both sides.

\begin{example}
To demonstrate the limitations of LRD, we translate rule 1 (written as one FRD rule) into a mapping containing only LRD rules,
\[
\forall i, g \;
  \left(
    \text{meds}(i, g) \rightarrow
    \exists x_1, x_2, x_3, x_4 \;
      \text{Drugs}(x_1, \tau_1[g], x_2, x_3, x_4)
  \right)
\]
\begin{equation*}
\begin{aligned}
\forall i, f, m, d, y \; \big(&
    \text{trial}(i,f, m, d, y), y>1990 \\
    &\rightarrow
    \exists x_5 \;
      \text{Clinical\_Trials}(x_5, f, \tau_2[m, d, y])
  \big)
\end{aligned}
\label{eq:gav}
\end{equation*}
Note that the translation from FRD to LRD requires two rules, isolating the source and target relations.
This is problematic since the variables $i$ and $x_1$ do not share the same scope across rules.
In other words, the LRD mapping will result in a target instance which does not specify which clinical trials concern which drugs.
Further, the instance will contain all drugs in $meds$ regardless of their most recent clinical trial.
\end{example}

\paragraph{Schema Alignments}
Rather than generating schema mappings directly, many works focus on the simpler task of generating schema alignments, which can eliminate many undesired mappings from consideration.
A schema alignment is a set of pairs, 
$\{(s_l, g_k) \mid s_l \in attr(S_i), g_k \in attr(G_j)\}$
where a pair asserts that source attribute $s_l$ semantically corresponds to target attribute $g_k$.
In limited cases, algorithms can produce the exact correct mapping rules given the alignments as input \cite{popa2002translating}.
Figure~\ref{fig:connecting} includes schema alignments for our running example.

\subsection{Problem Definition}
Given a source schema $\mathcal{S}$, a global schema $\mathcal{G}$, and a set of \textit{hints}, a \textit{Schema Mapping Assistant} must produce a candidate set of mappings $C$ of some class as described in Section~\ref{subsec:expressivness}.
From the schemata itself, we are guaranteed to have certain information, including relation names, attribute names, and any constraints stated within the schema (e.g., primary keys, foreign keys, data types, etc,.).

\subsubsection{Hints}
In addition to the information given by the schemata, we may also have additional contextual information, which we call \textit{hints}, that can be leveraged for determining $C$.
In current work, we consider natural language descriptions of tables $tdesc(\cdot)$ and attributes $adesc(\cdot)$, as well as sample data values $val(\cdot)$.
\begin{example}
For example, the \textit{Drugs} in Figure~\ref{fig:connecting} might have the following hints,
\begin{align*}
    name(T) & = "Drugs"\\
    tdesc(T) &= "Information\ on\ class,\ uses,\ and\ side\ effects"\\
    datatype(T) & = \{int, String, String ..., \}\\
    adesc(T) & = \{"Unique\ ID\ for\ Drug.",\\
    & "Brand\ name\ of\ drug.", ..., \}
\end{align*}
Not pictured are other hints that are given by the schema definition.
\end{example}

The availability of additional hints depends on the sources themselves.
Data values may not be available, either because the relations are empty or the data itself is restricted due to privacy concerns.
In practice, natural language descriptions may be derived from documentation, but even this is not guaranteed to exist.

\subsubsection{Candidate Set Quality}
The goal is to provide users with a high-quality candidate set.
Ultimately, what makes a candidate set "high-quality" depends on many factors, including user preference.
However, we often prioritize recall over precision:  
 maximize the number of true mappings while simultaneously minimizing the number of false mappings.
This is because it often takes less effort for a human to confirm that a candidate mapping is wrong than it does for them to determine the correct mapping \textit{on their own}.
\section{Challenge: Inconsistent Outputs}
\label{sec:inconsistent_outputs}

The output of LLMs depend heavily on how their input is phrased. 
The space of possible outputs is often diverse and varies significantly in terms of quality.
Further, it is difficult to predict the relative impact that different phrasings will have on the output. 
For example, minor adjustments in the ordering of content—such as rearranging rows or columns in a table—can impact accuracy, as LLMs are sensitive to the structure of the input data \cite{TableMeetsLLM}.

However, most existing schema matching approaches overlook these sensitivities. They typically rely on a single, static prompt and do not account for structural variations in the input. For example, Parciak et al \cite{TaDa} repeat the same prompt multiple times and aggregate the outputs using majority voting to approximate a high-confidence result, rather than varying the prompt itself. In contrast, we treat prompt variation as a key mechanism for exploring the model’s output space more effectively and improving overall quality of the final candidate set.

\subsection{Sampling Outputs}

Instead of relying on a single prompt, we treat this as a sampling problem. More specifically, we view the model as a black-box from which we can sample mappings by providing different prompt phrasings. More formally, we start with a reasonably effective prompt template $P(\cdot)$, and then apply a set of transformations that exploit symmetric properties of our problem. 
We prompt the LLM $n$ times to produce candidate sets $c_1, c_2, ... c_n$. 
We then combine these sets using a function $A(\cdot)$ to produce the final candidate set $C'=A(c_1, c_2, ... c_n)$. 

Sampling multiple outputs for the same input can benefit us in two key ways. First, it expands our coverage of the hypothesis space, increasing the chance of discovering high-quality alignments that might otherwise be missed with a single static prompt. Second, it provides insight into the relative likelihood of different mappings: if a particular output appears consistently across multiple samples, it may serve as a proxy for confidence in its quality.
 
\subsubsection{Symmetric Transformations}
Starting from a well-performing base prompt, we apply transformations that alter the input format without changing the underlying task. Specifically, we introduce variation by randomly permuting the order of columns in the prompt, sampling different data instances, and swapping the source and target tables. These transformations preserve the semantic equivalence of the task while encouraging the model to explore different parts of the output space.

\subsubsection{Sample Merging}
To combine the candidates generated from each prompt group (e.g., prompts using the original table orientation or the swapped orientation), we apply aggregation functions such as union, majority vote, or intersection. The choice of function depends on the desired trade-off between recall and precision. For example, union maximizes recall by including all possible matches, majority vote refines the output by selecting the most frequent alignments, and intersection focuses on the most consistent matches across all outputs.

Chen et al. \cite{consistency} explore LLM-based consistency sampling by prompting the model to select the most coherent response among a set of candidates. While their approach is designed for free-form generation, applying it directly to schema matching would require an additional LLM call over a long, concatenated prompt, which can be problematic for structured outputs with high token length. Moreover, selecting a single response is incompatible with our setting, where multiple candidate mappings per attribute must be considered and further analyzed. Instead, we apply logical operations such as majority vote over structured JSON outputs, providing a lightweight and scalable alternative without added inference cost or context limitations.

\subsection{Bidirectional Schema Matching}

To account for alignments from both the original and swapped table perspectives, simple aggregation functions like majority vote or union are not ideal. These methods fail to consider the differing confidence levels that the LLM may have for matches from each perspective. Instead, we propose estimating a confidence score for each candidate match in both directions before merging the results, leading to a more reliable combination.

MatchMaker estimates the LLM's confidence in alignments by prompting it to score candidate matches for each attribute. However, these scores may not accurately reflect true probabilities. Inspired by \cite{wang_logits}, we approximate the LLM’s confidence by first asking it to select the best match among the candidates, then using its output logits to compute confidence scores.

Combining results from the original and swapped table perspectives presents a challenge, as its not clear which merging method is most effective. We explore different techniques for this task. As a starting point, we test two simple approaches: averaging and multiplying confidence scores. Averaging reduces the confidence if an alignment appears in only one direction, while multiplication removes alignments that are missing from either direction. 

In addition, we apply the stable matching algorithm \cite{stable_matching} to the ranked matches from both directions. This approach conceptually aligns with the table-swapping process, as it reflects the bidirectional nature of preferences and ensures a stable and consistent matching between attributes. We define Stable Schema Matching as follows:

\begin{definition}[Stable Schema Matching]
    Given two schemas \(A\) and \(B\), the input consists of two sets of attributes, \(A[\text{attr}] = \{a_1, a_2, \dots, a_n\}\) and \(B[\text{attr}] = \{b_1, b_2, \dots, b_m\}\), along with ranked preference lists for each attribute in \(A[\text{attr}]\) over the attributes in \(B[\text{attr}]\), and vice versa. The goal is to compute a stable matching between attributes in \(A\) and \(B\) such that no unmatched pair of attributes would prefer each other over their current matches. The output is a set \(M = \{(a_i, b_j) \mid a_i \in A[\text{attr}], b_j \in B[\text{attr}]\}\) containing the stable matches between attributes in \(A\) and \(B\), ensuring that the matching is mutually acceptable and stable. The number of matches can be constrained by a parameter \(K\), which specifies the top \(K\) stable matches for each attribute.
\end{definition}

\begin{figure}[H]
  \centering
  \includegraphics[width=\columnwidth]{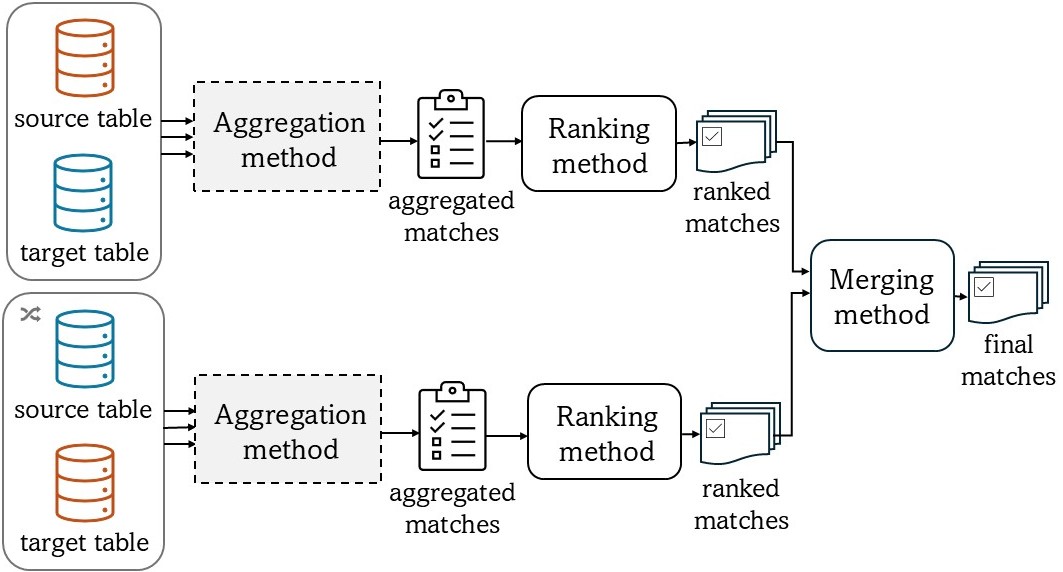}
  \caption{Bidirectional schema matching process. Matching is performed in both directions by swapping the roles of source and target tables. Each direction involves aggregation and ranking of candidate matches, which are then merged to produce the final results.}

  \label{fig:sym}
  \Description{
  A diagram with two parallel pipelines for schema matching. The top pipeline starts with a source and a target table, runs them through an aggregation method, then a ranking method, and produces ranked matches. 
  The bottom pipeline performs the same steps but swaps the roles of source and target. 
  The outputs of both directions are merged to generate the final matches.
  }
\end{figure}

 An overview of our bidirectional matching design is shown in Figure~\ref{fig:sym}. For more details on our prompts, how we compute the confidence score, and the stable matching algorithm, please refer to the technical report\footnote{The technical report can be found at \url{\techreportURL}}.

\subsection{Preliminary Results}
\label{SM-results}
In this section, we evaluate the impact of our proposed prompting strategies and bidirectional approach with symmetric transformations, demonstrating how these methods help an open-source model achieve competitive performance compared to proprietary models.

\paragraph{Problem Definition} We focus on the simplest form of mapping, which is nonetheless a difficult task.
Specifically, given relations $S_i$ and $G_j$, we want to to produce a set of pairs, each representing an alignment between a source attribute and a global attribute. 
Formally,  $C=\{(s_l, g_k) \mid s_l \in attr(S_i), g_k \in attr(G_j)\}$ indicating that the attribute $s_l$ semantically corresponds to the attribute $g_k$.

\paragraph{Prompt Template and Techniques.}
We use reasoning-based prompting strategies to improve LLM performance in zero-shot settings \cite{zeroshotReasoner}. 
Specifically, we prompt the LLM three times with transformed inputs using a fixed seed. 
To ensure consistent output formatting and enable efficient processing, we constrain the model to produce structured JSON output \cite{prompt_conf}.
Each prompt follows an $N$-1 format, where $N$ source attributes and one target attribute are serialized in JSON to improve structural understanding \cite{TableMeetsLLM, TabularRepr}.
This $N$-1 strategy has been shown to outperform other settings in terms of matching effectiveness \cite{TaDa}. We conducted experiments to evaluate the impact of different types of metadata. Our results show that natural language descriptions of attributes consistently provide the largest boost to schema matching performance. Even with basic schema details such as table names, attribute names, and data types, the LLM-based method outperforms traditional approaches such as COMA in the single-prompt setting. We included data values based on the idea that example values could help the model better understand attribute meaning. Interestingly, data values caused a slight drop in performance in single-prompt setups but proved more helpful when aggregating results across multiple prompts—each varying in column order and sampled values. 
For more details and results, please refer to our technical report. To improve the semantic reasoning of the LLM, we include all available schema metadata in the prompt, including attribute names, data types, descriptions, ten randomly sampled unique data values, and relation descriptions.

\paragraph{Datasets}
We conduct our experiments on two widely used schema matching datasets: MIMIC-OMOP and Synthea-OMOP.
To populate OMOP with data values, we used sample data from MIMIC-IV formatted under the OMOP model. MIMIC-IV differs enough from MIMIC-III to avoid record overlap but remains conceptually similar. However, some aligned columns lack sufficient sample data due to privacy restrictions or missing data. Many columns also lack descriptions, which adds complexity to schema matching.
MIMIC-OMOP includes 26 schema pairs that align real-world healthcare databases: MIMIC-III and the OMOP Common Data Model. It covers 268 source attributes, 203 target attributes, and 155 ground truth matches. This dataset reflects the complexity of real-world medical data. 
In contrast, Synthea-OMOP is based on synthetic healthcare records generated by Synthea and aligned to OMOP. It contains 12 schema pairs with 101 source attributes, 134 target attributes, and 105 matches. Although synthetic, Synthea captures realistic data variability and schema ambiguity. This makes it a challenging and widely used benchmark for schema matching tasks.

\paragraph{Baselines}
We compare our methods with three baselines that require no training data. The first is COMA \cite{Coma1}, a widely used schema-matching method known for its efficiency and flexibility, which has been refined through several iterations \cite{Coma++, Coma3}. COMA is a rule-based approach that lacks semantic understanding, with the schema-based version considering only schematic information, and the instance-based version incorporating both schematic information and data values. As a result, COMA can miss important semantic nuances and struggle with the complexities of real-world schemas like those in MIMIC-OMOP and Synthea-OMOP, which affects its overall matching performance. We evaluate both versions using Valentine’s Python wrapper for COMA 3.0\footnote{\url{https://github.com/delftdata/valentine}} \cite{valentine}.

For language model baselines, we use the N-1 prompting method from Parciak et al. \cite{TaDa}. This method is simple and effective. It works by aggregating multiple prompts, similar to our own aggregation approach. However, It does not provide ranked match suggestions, which are often useful in practice. It also restricts mappings to one-to-one correspondences, overlooking scenarios where an attribute may align with multiple counterparts.
We also evaluate MatchMaker \cite{MatchMaker}, which refines alignments with pre- and post-filtering steps. Although MatchMaker improves performance by filtering, its multi-step prompting pipeline introduces inefficiencies and can fail if intermediate language model outputs deviate from expected patterns. Both methods are implemented based on their respective papers, and for a fair comparison, we exclude the pre-filtering step in MatchMaker, as our method does not include it

\paragraph{Metrics}
We report the average Precision@k, Recall@k, and F1@k for both our approach and the LLM baselines, averaged across three different random seeds. For methods that do not rank the alignments, we report metrics at $k = \text{max}$. For the bidirectional methods using stable matching and multiplication, the value of k is limited based on the pipeline, as the final alignments must be present in the aggregated candidates from both directions. Therefore, we report a limited k for these methods. For the method using averaging, we report the maximum k obtained from the MatchMaker method.
In real-world data integration scenarios, automated matching candidates require manual validation. Therefore, we aim to achieve high precision and recall simultaneously to reduce manual effort while ensuring high-quality matches.

Many studies across various tasks, including schema matching, demonstrate that larger language models perform better \cite{TaDa, zeroshotReasoner, largelanguagemodelsdata, selfConsistency}. 
However, the most advanced models are typically available only via APIs, which raises significant privacy concerns.
In these experiments, we use the Meta-\textit{Llama-3.1-70B-Instruct-GPTQ-INT4}
\footnote{The "70B" refers to the model's size, with 70 billion parameters. "Instruct" indicates that the model is fine-tuned for instruction-following tasks. "GPTQ" is a quantization method that optimizes memory efficiency and improves inference speed. "INT4" refers to 4-bit integer quantization, a specific technique used within GPTQ to further reduce memory usage while maintaining performance.}
, the largest model we could run on the server. More details on the evaluation setup can be found in our technical report.

\subsubsection{\textbf{Evaluation Results}}

\begin{table}[hb]
\small
\hspace*{-0.06in}
\begin{tabular}{|l|l|c|c|c|c|}
\hline
\textbf{Method} & \textbf{Experiment} & \textbf{k} & \textbf{P@k} & \textbf{R@k} & \textbf{F1@k} \\
\hline
\multirow{2}{*}{Aggregation} & Original tables& max & 0.35 & 0.79 & 0.47 \\
                             & Swapped tables& max & 0.47 & 0.67 & 0.54 \\
\hline
\multirow{6}{*}{Bidirectional} & \multirow{2}{*}{Stable Matching} & 1 & 0.68 & 0.62 & 0.64 \\
                               &                     & 2 & 0.66 & 0.63 & 0.64 \\
\cline{2-6}
                               & \multirow{3}{*}{Average} & 1 & 0.37 & 0.78 & 0.49 \\
                               &                      & 2 & 0.31 & 0.82 & 0.44 \\
                               &                      & 3 & 0.30 & 0.83 & 0.43 \\
\cline{2-6}
                               & \multirow{2}{*}{Multiply} & 1 & 0.67 & 0.62 & 0.64 \\
                               &                            & 2 & 0.66 & 0.63 & 0.64 \\
\hline
\multirow{6}{*}{Baseline} & \multirow{3}{*}{MatchMaker}& 1 & 0.25 & 0.24 & 0.23 \\
                          &  & 2 & 0.15 & 0.30 & 0.19 \\
                          & & 3 & 0.11 & 0.31 & 0.15 \\ 
\cline{2-6}

                          & \multirow{1}{*}{COMA Sch.} & max & 0.14 & 0.11 & 0.10 \\
                          & \multirow{1}{*}{COMA Inst.} & max & 0.21 & 0.14 & 0.16 \\
\cline{2-6}
                          & \multirow{1}{*}{Parciak et al} & max & 0.30 & 0.15 & 0.18 \\
\hline
\end{tabular}
\caption{Precision@k (P@k), Recall@k (R@k), and F1@k for different methods on the MIMIC dataset. 
}

\label{tab:bidir-ehr}
\end{table}

\begin{table}[hb]
\small
\hspace*{-0.06in}
\begin{tabular}{|l|l|c|c|c|c|}
\hline
\textbf{Method} & \textbf{Experiment} & \textbf{k} & \textbf{P@K} & \textbf{R@K} & \textbf{F1@K} \\
\hline
\multirow{2}{*}{Aggregation} & \multirow{1}{*}{Original tables} & max & 0.57 & 0.95 & 0.70 \\
                             & \multirow{1}{*}{Swapped tables}  & max & 0.52 & 0.56 & 0.51 \\
\hline
\multirow{6}{*}{Bidirectional} & \multirow{2}{*}{Stable Matching} & 1 & 0.78 & 0.52 & 0.60 \\
                               &                     & 2 & 0.77 & 0.55 & 0.62 \\
\cline{2-6}
                               & \multirow{3}{*}{Average} & 1 & 0.58 & 0.95 & 0.71 \\
                               &                      & 2 & 0.50 & 0.95 & 0.65 \\
                               &                      & 3 & 0.48 & 0.96 & 0.63 \\
\cline{2-6}
                               & \multirow{2}{*}{Multiply} & 1 & 0.77 & 0.55 & 0.62 \\
                                & & 2 & 0.77 & 0.55 & 0.62 \\
\hline
\multirow{9}{*}{Baseline} & \multirow{3}{*}{MatchMaker} & 1 & 0.45 & 0.21 & 0.27 \\
                          &                              & 2 & 0.23 & 0.21 & 0.21 \\
                          &                              & 3 & 0.15 & 0.21 & 0.17 \\

\cline{2-6}
                          & \multirow{1}{*}{COMA Sch.} & max & 0.30 & 0.15 & 0.19 \\ 
                          & \multirow{1}{*}{COMA Inst.} & max & 0.30 & 0.16 & 0.20 \\ 
\cline{2-6}
                          & \multirow{1}{*}{Parciak et al} & max & 0.53 & 0.11 & 0.17 \\
\hline
\end{tabular}
\caption{Precision@k (P@k), Recall@k (R@k), and F1@k for different methods on the Synthea dataset.}

\label{tab:bidir-synthea}
\end{table}

The results in this paper use the majority vote aggregation method, as it offers the best balance between recall and precision. Results for other aggregation methods are available in our technical report.
As shown in Tables ~\ref{tab:bidir-ehr} and \ref{tab:bidir-synthea}, our aggregation method outperforms all baselines. 
In addition to our symmetric transformations, the key difference from Parciak et al. lies in the output format and schema serialization used in the prompt. Their method struggles with incomplete responses, where the model often skips the final decision\footnote{Example: \textit{The first attribute to consider is $\{source\_attribute\}$. Does $\{source\_attribute\}$ semantically match $\{target\_attribute\}$?}}. 
MatchMaker also fails when the LLM does not follow the required format in intermediate steps.
While both MatchMaker and Parciak et al. used GPT-4 in their studies, we employed a significantly smaller model in our experiments. 
As a result, our prompt is more effective for smaller models, which we attribute to its clear output format and structured JSON schema serialization.

On the MIMIC dataset, LLM-based methods outperform COMA, highlighting that LLMs excel in domains where domain knowledge is crucial, thanks to their ability to process natural language descriptions. In contrast, on the Synthea dataset—where vocabulary and attribute names require less domain knowledge—Parciak et al.'s method does not outperform COMA. This suggests that when domain knowledge is less critical, the choice of pipeline and prompting method, especially for smaller models, becomes more important.

We also observe that the bidirectional method using multiplication outperforms the others, achieving the best F1@1 score. The bidirectional method using stable matching closely follows. The difference lies in our use of confidence score ranking in stable matching, while multiplication computes final alignment confidence based on the actual values of scores from each direction. These methods are effective for tasks where both precision and recall are important. The bidirectional method using averaging, though lower in precision, excels in recall and is preferred when high recall is prioritized.

We compared our methods against LLM baselines originally designed for GPT-4. MatchMaker is directly comparable, as it also evaluates on the MIMIC and Synthea datasets. We assess performance by comparing our bidirectional method, based on an open-source model, with the full MatchMaker pipeline using their reported accuracy@1. As shown in Table~\ref{tab:vs-gpt4}, our bidirectional method using stable matching and multiplication achieves accuracy comparable to MatchMaker on the Synthea dataset. It also outperforms MatchMaker on the MIMIC dataset.

\begin{table}[ht]
\centering
\small
\begin{tabular}{|c|l|c|}
\hline
\textbf{Dataset} & \textbf{Method} & \textbf{Accuracy@1} \\
\hline
\multirow{5}{*}{MIMIC} & MatchMaker & 62.20 $\pm$ 2.40 \\

\cline{2-3}
& Bidirectional (Stable Matching) & 0.78 $\pm$ 0.00 \\
& Bidirectional (Average) & 0.49 $\pm$ 0.01 \\
& Bidirectional (Multiply) & 0.77 $\pm$ 0.01 \\
\hline
\multirow{5}{*}{Synthea} & MatchMaker & 70.20 $\pm$ 1.70 \\

\cline{2-3}
& Bidirectional (Stable Matching) & 0.69 $\pm$ 0.01 \\
& Bidirectional (Average) & 0.64 $\pm$ 0.01 \\
& Bidirectional (Multiply) & 0.70 $\pm$ 0.01 \\

\hline
\end{tabular}
\caption{\small Comparison of our proposed method using Llama-3.1-70B-Instruct-GPTQ-INT4 against MatchMaker's using GPT-4.}
\label{tab:vs-gpt4}
\end{table}

Our bidirectional approach, combined with symmetric transformations, delivers strong results on clinical datasets like Synthea and MIMIC, achieving accuracy comparable to or exceeding GPT-4-based models such as MatchMaker. While their results were obtained using a significantly larger model, our approach, built on a smaller open-source model, performs competitively. This highlights that strong performance can be achieved not just through model scale or domain knowledge, but through careful pipeline design. In addition to improved prompting strategies, such as more effective schema serialization, our method introduces order variations at the table, column, and data value levels to encourage broader exploration of the alignment space. These design choices highlight the impact of pipeline structure in maximizing the effectiveness of LLMs for schema matching, even without access to proprietary models.

\subsubsection{Sampling Techniques for FRD Mappings}

Our evaluation indicates that, with the correct sampling techniques, open source LLMs are highly competitive with at least one major, proprietary model (GPT-4).
However, these techniques depend on decomposing the LLM's responses into a fairly limited number of atomic elements.
For example, the final answer in the LLM's response for schema alignment is pairs of source columns and target columns.
It is easy to break the output into sets of these pairs, making the ordering irrelevant,
Further, each pair is reflexive in the sense that (A, B) is identical to (B, A), allowing us to leverage bidirectional techniques.
In essence, sampling techniques require outputs that are simple enough such that they can be broken down into atomic elements and overlap between responses can be calculated.

Extending these techniques to schema mappings is not as easy do to the increased complexity in the output language.
A mapping is essentially a collection of queries, and it is not immediately clear how one might effectively partition such an output to test for overlap.
One way to partition the output is on a per-query basis, testing for the repetition of queries within outputs.
However, using exact string comparison would be much too restrictive, likely leading to very little overlap, even if there are logically-equivalent queries.
Simultaneously, the logical-equivalency of queries is not guaranteed to be decidable \cite{abiteboul1995foundations}.
Further, if a given rule is not invertible, then the bidirectional technique cannot be used.
This makes such techniques incorrect in a theoretical sense, but empirically, such techniques may still be useful if employed correctly.
Clearly, this is an important subject for future research.

\section{Challenge: Representation and Large Input}
\label{sec:complex_mappings}

As discussed in Section~\ref{subsec:expressivness}, existing research does not consider FRD rules making the associated techniques insufficient for many common mapping scenarios.
In this section, we consider two immediate challenges associated with Zero-Shot generation of FRD mappings using LLMs.
First, we discuss the difficulties with representing FRD rules.
Second, we discuss the overall complexity of generating a full FRD mapping.
We close this section with an empirical study meant to help us better understand these challenges and establish future research.

\subsection{Output Representation}

LLMs have show great success at generating SQL \cite{liu2024survey}, making it a promising candidate for representing rules.
However, an SQL query can only represent a single GaV rule because it's output is always only a single table.
In fact, most popular query languages can only produce individual GaV rules.
That being said, multiple SQL statements can be used in tandem to represent a GLaV rule.
Whether an LLM will produce such scripts is another issue. 
That being said, it is possible to translate SQL into other classes.
For example, in the context of importing tabular data into a relational target, one work prompts an LLM to generate a query over the target and then inverts it to produce a LaV rule \cite{huang2150transform}.
However, this approach only works if the GaV counterpart is invertible itself, a theoretically needless limitation.

\subsection{Input}
Given some unknown rule, the LLM can only generate it if it is provided, at minimum, the schema fragments appearing in the rule.
On the other head, existing works have shown that filtering input to that which is relevant often improves performance.
It is reasonable to believe that we would also benefit from filtering out irrelevant schema.
However, such a task is not straightforward.

A mapping could be quite large, with rules covering the entirety of the source and target schemas.
In such an instance, it perhaps makes more sense to break the full mapping down into parts.
A sensible approach would be to segment the full mapping at the rule level.
However, since we do not know the underlying rules, we would need to predict their contents related to source and target relations.

\paragraph{Relations within Rule k}
As the same relation may appear multiple times within a conjunction, we define $\mathcal{S}^k$ and $\mathcal{T}^k$ to be the unique relations appearing in $\phi(\vec{x})$ (i.e., source relations appearing in left-hand side of the rule) and $\psi(\vec{x}, \vec{y})$ (i.e., target relations appearing in right-hand side of the rule), respectively, for the kth rule.

We do not know either $\mathcal{S}^k$ or $\mathcal{T}^k$ for any given rule (much less the rule itself).
And finding $\mathcal{S}^k$ and $\mathcal{T}^k$ is, itself, a form of schema filtering.
However, knowing neither set of relations makes filtering a potentially multi-step process where we must predict subgraphs of the target and the source and then pair those subgraphs across the source and target correctly.
This may be quite hard, perhaps motivating solutions that are very relaxed in how they filter schema.

When developing a schema filter, we must consider its restrictiveness: we want a filter that maintains as many relevant schema parts as possible while also reducing the size of the context.
This presents an optimization problem where we must decide how aggressively to filter such that the LLM's performance does not significantly degrade either due to a lack of context (information) or an overly noisy context (too much irrelevant information).

\subsection{Preliminary Results}
In this section, we present experiments meant to provide some preliminary insight into the two questions raised in this section: 1) in what ways is SQL sufficient (insufficient) for generating GLaV rules? and 2) How does the input size affect the final set of mappings?
For the later question, we assume that we have already found $\mathcal{S}^k$ and $\mathcal{T}^k$ for each kth rule.
The question is, how do we chunk these specifications into prompts.

\paragraph{Prompt Template and Techniques}
Similar to our prompt discussed in Section~\ref{SM-results}, we assume a zero-shot setting and encourage the model to reason about its mapping prior to producing its final script.
Each prompt can contain multiple source relations and target relations, each of which is serialized as JSON.
We include all available and relevant metadata.
For each relation, we include its name, primary key, and any foreign key relationships.
For each attribute, we include its name, type, whether it is NULLable, and up to 10 instance data values, uniformly sampled without replacement.
Each sampled data value is truncated to 100 characters.

\paragraph{Dataset}
We use a subset of Amalgam \cite{Mil+01}, a commonly-cited schema mapping benchmark.
Of Amalgam's four bibliography databases, we take S1 as our source and S2 as our target.
The gold mapping contains 7 rules.
Generally speaking, it involves decomposing publication-type relations (S1) into attribute-specific relations (S2).
During our research, we have noticed a surprising lack of schema mapping benchmarks despite the prevalence of the problem.
Unfortunately, many benchmarks have broken links.
We have pieced Amalgam together from a few places:
we use the schema definitions from the original source \cite{Mil+01}, the data provided by \cite{kruse2015estimating}, and the ground truth as provided by the iBench scenario Github page\footnote{\url{https://github.com/RJMillerLab/ibenchScenarioCollection}}.
Notably, the ground truth mappings are specified using a proprietary format.
We translate these to SQL for our purposes.
We hope that the inaccessibility of the original dataset helps circumvent any data leakage issues.
Amalgam does not contain any natural language descriptions of attributes or relations.
However, given its common domain, we hypothesize that an LLM should have a general understanding of the domain.

\paragraph{Metrics}
We borrow a metric commonly used in Text-to-SQL called execution accuracy which measures the overlap between the results obtained from predicted queries and ground-truth queries.
However, instead of only reporting either 1 (full overlap) or 0 (anything less than full overlap), we report the percentage of overlap.
As discussed in Section~\ref{sec:schema_mapping}, most data integration systems follow a human-in-the-loop approach, where a user can validate and fix predicted mappings.
Thus, it is useful to report the proximity of a predicted mapping to that of the ground truth.

To test overlap, we need to compare the effect of both mappings given the same input target-instance.
For evaluation data, we generate 100 rows for each table in the target database.
When schematically valid, we insert NULLs into attributes for some rows and remove some foreign key references, resulting in some parent rows with no children (i.e., some authors with no publications and some publications with no authors).
We apply both the gold and predicted mapping to produce a gold target instance $I'$  and a predicted target instance $I$.
We then apply an exhaustive set of test queries to both instances.
For each test query $q_i$, we calculate false positive rows (FP), false negative rows (FN), and true positive rows (TP) as follows,
\begin{equation*}
FP_i = q_i[I] - q_i[I']
\qquad
FN_i = q_i[I'] - q_i[I]
\qquad
TP_i = q_i[I'] \cap q_i[I] 
\end{equation*}
We then calculate recall (R), precision (P) as,
\begin{equation*}
R_i = |TP_i| \ / \ (|TP_i| + |FN_i|)
\qquad
P_i = |TP_i| \ / \ (|TP_i| + |FP_i|)\\
\end{equation*}
Finally, F1-score is calculated as $F1_i = (2.0 * R_i * P_i) \ / \ (R_i + P_i)$.

We test for two kinds of overlap: Table Overlap and Join Overlap.
In both cases, we project over all columns \textit{except for arbitrary primary keys and foreign key references}.
For our dataset, these keys have no semantic meaning.
Thus, correctness depends on upon establishing the correct references between rows and not on the particular values used to do so. 
our only expectation is that, no matter the values assigned to these keys, the correct references are established between rows.

\paragraph{Table Overlap}
Let $T_1(x_1, \bar{x}_1), T_2(x_2, \bar{x}_2), ..., T_n(x_n, \bar{x}_n)$ be relations in the target schema, where $x_1, x_2, ..., x_n$ is each relation’s respective set of attributes (columns), and $\bar{x}_1, \bar{x}_2, ..., \bar{x}_n$ are the primary key columns and foreign key references for their respective relation.
For each $T_i$, we test for overlap using $q_i(x_i) :- T_i(x_i \cap \bar{x}_i)$ if both $q_i[I']$ and $q_i[I]$ return no results, then we do not consider $q_i$ in our final calculation.
This prevents the overlap from being inflated by target tables that are not touched by any mapping.

\paragraph{Join Overlap}
We indicate the target relations appearing in the kth gold mapping with $\mathcal{T}_k = T_1(x_1, \bar{x}_1), T_2(x_2, \bar{x}_2), ..., T_l(x_l, \bar{x}_l)$.
For each $\mathcal{T}_k$, we test for overlap using $q_k(\bigcup_{j=1}^{l} x_j) :- T_1(x_1 \cup \bar{x}_1) \bowtie T_2(x_2 \cup \bar{x}_2) \bowtie ... \bowtie T_l(x_l \cup \bar{x}_l)$.
In instances where $|\mathcal{T}_k| = 1$, we drop query $q_k$ as it devolves to a query with no joins, which is already covered by Table Overlap.
In instances where we have multiple, identical queries due to target-relation overlap in our gold mapping, we remove all but one of the queries.
After calculating individual metrics, we take the average of all queries.

It is worth noting that perfect Join Overlap (i.e., F1 = 1.0) also implies perfect Table Overlap, but perfect Table Overlap does not imply perfect Join Overlap.
In fact, low Table Overlap is likely connected with very low table overlap scores given the nature of the metrics.
Producing high Join Overlap requires that the model effectively generate the correct references between rows, sometimes requiring the model to invent new keys.

\subsubsection{Evaluation Results}
We conduct an empirical evaluation to answer two questions.
First, we want to know how does chunking affect performance; namely, are there benefits of producing more chunks at a time. What are the drawbacks of having the LLM do more work with fewer prompts?
Second, we want to better understand the limitations of a baseline approach, which will help establish promising research directions.
\begin{figure*}
    \centering
    \includegraphics[width=1\linewidth]{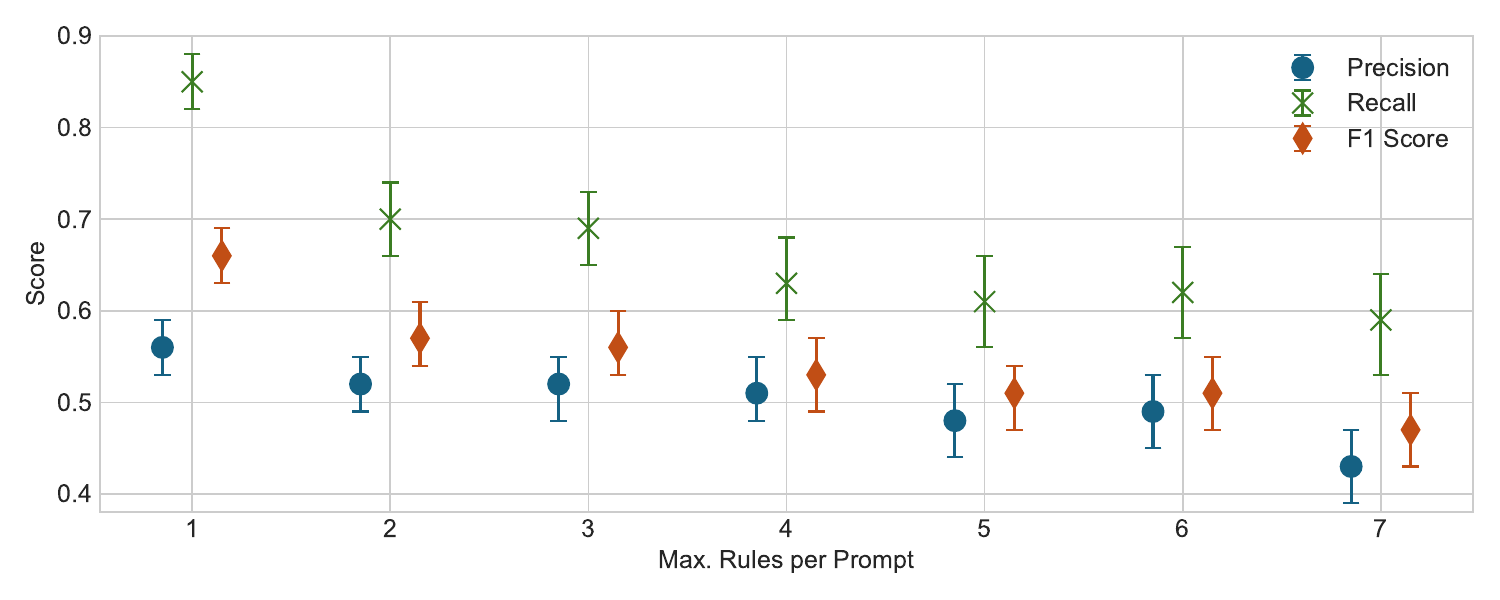}
    \caption{Average precision, recall, and F1-Score plotted against the maximum rules per prompt (MRPP). 
    Error bars represent a 95\% confidence interval.}
    \label{fig:rules_per_prompt}
    \Description{Average precision, recall, and F1-Score plotted against the maximum rules per prompt (MRPP). 
    Error bars represent a 95\% confidence interval.}
    
\end{figure*}

\begin{table}[ht]
\centering
\small
\begin{tabular}{|c|c|c|}
\hline
\textbf{Rules / Prompt} & \textbf{Input Tokens} & \textbf{Output Tokens} \\
\hline
1 & 3910 & 1104 \\
\hline
2 & 5024 & 1484 \\
\hline
3 & 6425 & 1838 \\
\hline
4 & 7596 & 2053 \\
\hline
5 & 8839 & 2489 \\
\hline
6 & 9983 & 2849 \\
\hline
7 & 11259 & 2639 \\
\hline
\end{tabular}
\caption{\small The average number of input and output tokens according to the number of rules a prompt is given specification for (i.e., the underlying source and target relations).
Each average is calculated over 20 random prompt/response pairs.
}
\label{tab:avg-tokens-by-rules_per_prompt}
\end{table}

To understand the effect of input size on mapping quality, we vary the input size of prompts with respect to how many $(\mathcal{S}^k, \mathcal{T}^k)$ pairs we supply in the prompt.
As more pairs are added, the model will not only need to parse more information, but will also need to generate more code.
We treat the \textit{Max. Rules per Prompt (MRPP)} as a hyperparameter and vary it from 1 to 7.
For example, if MRPP is 5, we will prompt the model twice: once with a specification for 5 rules and again with a specification for 2 rules (one prompt will have fewer rules if the number of mappings is not divisible by MRPP).
Rules are uniformly sampled without replacement, and as we add rules, we remove overlapping relations.

We use the same LLM here (Meta-Llama-3.1-70B-Instruct-GPTQ-INT4).
For each MRPP, we prompt the model 20 times using different seeds, controlling the order in which relations and attributes within those relations are presented and which data values are sampled.
Unlike the previous section, we do not combine the outputs of these different models, but rather use this as a way to get a more robust measurement of performance that is not dependent on a single representation of the input.
As discussed, combining these outputs is a viable technique for producing better programs, but it is saved for future works.
We report 95\% confidence intervals for our metrics.

Figure~\ref{fig:rules_per_prompt} shows the performance averaged over seeds for each setting for MRPP. 
Though we see a decrease in performance overall, it is most drastic for recall, implying that the model omits parts the rules whenever more than one rule is included in the prompt.
Table~\ref{tab:avg-tokens-by-rules_per_prompt} gives us an idea of how the complexity of the input and output grows as more rules are added.
\section{Challenge: Reducing Poor-Quality Mappings}
\label{sec:mapping_quality}
In Section~\ref{sec:inconsistent_outputs}, we proposed methods for filtering the candidate set based on sampling multiple outputs from an LLM.
Furthermore, we empirically showed how such methods can improve the candidate set of mappings.
However, sample-based filtering will not eliminate false positives if they consistently appear in the output.
Thus, it is worth considering other filtering strategies that can complement these sample-based methods.
We propose additional methods for filtering the candidate set based on markers of mapping quality.

\subsection{Constraint-Based Filtering} 
Mappings must be consistent with both the explicit and implicit (semantic) constraints of the global schema. 
Explicit constraints include those stored in the schema and are often enforced by database management systems.
For example, mappings must populate required attributes and respect attribute data types. 
Moreover, mappings must respect semantic constraints, which are often not enforced by the database, but are still generally followed by its users. 
Some semantic constraints are commonsense regardless of the domain. 
For example, each row within a relation must have some information content. 
This constraint can be violated when a relation has an arbitrary primary key, which is a common database design practice. 
This can create situations where the primary key is populated but all of the attributes of the row are NULL, functionally producing a row with no real information content.

Another source of semantic constraints are business rules, which are often documented (likely as an entity-relationship diagram) as part of the global schema's design.
Some semantic constraints are commonsense given the domain (e.g., each child must have exactly two biological parents) while others may be esoteric and purely process-driven.
Regardless, many such constraints are often not stored in the schema.
Instead, they are very likely enforced by the end-user when they inspect the final mapping set.
In fact, the existence of latent constraints is one important reason for why humans \textit{must} validate candidate rules.

Specifying such constraints could be valuable in eliminating unfavorable mappings from the candidate set.
Further, since such constraints specify how data should be organized within the target schema, theoretically, they could be specified once and then be used to produce higher-quality candidate sets for \textit{all sources}, the very definition of a scalable technique.
Of course, it is likely that even these constraints will need to be edited as the target itself evolves.

Using semantic constraints is a promising technique for enabling scalable data integration systems, but there are important questions: how would one explicitly define these constraints and how could they be used in practice?
It would tedious and difficult to write out and manage all of these underlying constraints, so a more practical approach would be to derive them from other sources.
For example, if we already have data in our target, we can use techniques to derive constraints from that data.
Further, some such constraints could be learned through user preferences over candidate mappings.

\subsection{Model-Based Filtering}

Users may not want to semantic constraints into logical statements as it, admittedly, may require significant overhead.
Alternatively, one can leverage models of what consistent data "looks like". 
When adding new sources or maintaining existing ones, we can materialize the data (i.e., run the candidate mapping) and check the quality of the instance.

Assuming that the current instance data in the global up to this point is a good representation of "high-quality data." We can derive characteristics of that data that end up being indicators of quality. In some way, building a model that, through unsupervised learning, constructs its own non-logical black-box constraints of data quality based on the current instance. Reward models could essentially measure the similarity of of existing data and new data (proposed data from candidate mapping)--how well do they "mix".

Of course, this raises questions of how we might prevent the model from becoming biased towards our current instance. For example, if our current instance only contains publications from VLDB, the model might associate publication[title]="VLDB" as an important indicator of quality, and conversely, would assume that instances where, for some publications, publication[title]!="VLDB" indicates a poor mapping.
\section{Challenge: Efficient Prompting}
\label{sec:efficient_prompting}

\begin{table}[t]
\centering
\small
\begin{tabular}{|c|c|c|c|c|}
\hline
\textbf{Max. Rules/Prmpt} & \textbf{Input} & \textbf{Output} & \textbf{Total} & \textbf{Reduction} \\
\hline
1 & 27577 & 8135 & 35712 & --   \\
\hline
2 & 19204 & 5282 & 24486 & 1.46 \\
\hline
3 & 16635 & 4633 & 21268 & 1.68 \\
\hline
4 & 14087 & 3777 & 17864 & 2.00 \\
\hline
5 & 14083 & 4068 & 18151 & 1.97 \\
\hline
6 & 13884 & 4006 & 17890 & 2.00 \\
\hline
7 & 11259 & 2639 & 13898 & 2.57 \\
\hline
\end{tabular}
\caption{\small How chunking affects the number of tokens processed.
"Reduction" specifies efficiency relative the first row (i.e., using a separate prompt for each rule)}
\label{tab:avg-total-tokens-by-max_rules_per_prompt}
\end{table}

A major challenge in LLM-based schema mapping is its high computational cost: for large datasets, many tokens need to be processed.
Though computation is generally cheaper than human attention, it is necessary to consider the trade-off between performance and computational costs.
imprecisely speaking, techniques should produced the desired results without excessive computation. 
This is especially important for those who must pay a third party for computational resources.
We discuss three strategies for reducing computational costs (i.e., tokens processed).

\paragraph{Reducing Unnecessary Comparisons (Don't ask LLM trivial things)}

One way to reduce unnecessary comparisons is through data type prefiltering.
By categorizing attributes into broad types—such as Numeric, Text, Date/Time, and Boolean—we can reduce the number of source attributes that need to be compared to a target attribute. 
In an N-1 matching setup, where each prompt compares one target attribute to multiple source attributes, prefiltering ensures that only source attributes with the same data type as the target attribute are included in the comparison. 
This reduces the pool of source attributes from N to a smaller subset, k, improving both the efficiency and accuracy of the model by eliminating irrelevant comparisons.
Another approach is to use the results from the first round of predictions to assess confidence.
If the top match has a much higher score than the others, there may be no need for additional rounds of comparison, further reducing the number of calls.
By refining the selection process based on confidence and narrowing down the pool of candidate pairs, we can make the schema matching process more efficient and scalable.

\paragraph{Efficient Chunking}
As discussed, one tunable parameter is the amount of work induced (i.e., code written) by each prompt.
We observed in section~\ref{sec:complex_mappings} that the performance of our baseline approach does worsen as we ask it to generate more rules.
However, as shown in Table~\ref{tab:avg-total-tokens-by-max_rules_per_prompt}, we also observe that the total number of tokens processed (input/output) is drastically reduced when we have the LLM produce two rules instead of one.
Further, the total tokens processed is reduced by more than 50\% when we ask the model to generate the full mapping within one prompt.

The non-linear nature of this reduction implies that we can be more efficient with our techniques while also maintaining good performance--basically, the problem is more nuanced than simply reducing the number of prompts.
One reason for this non-linear relationship is that there is a constant overhead that must be paid with each prompt; namely, the static text which sets up our specific task for the LLM.
The other major reason has to do with high rule overlap, where underlying rules share many source and target relations.
Chunking two high-overlap rules together in one prompt has the affect of completing more work while adding relatively little to the input (i.e., the very few relations that do not appear in both rules).
In other words, some rules can be chunked together without adding significant complexity (i.e., number of tokens) to either the input or the output rule.
A simple technique, then, would be to chunk as many rules together under different prompts such that their overlap is high.

\bibliography{references}

\end{document}